# Imaging of Antiferroelectric Dark Modes in an Inverted Plasmonic Lattice

Javier Rodríguez-Álvarez,* Amílcar Labarta, Juan Carlos Idrobo, Rossana Dell'Anna, Alessandro Cian, Damiano Giubertoni, Xavier Borrisé, Albert Guerrero, Francesc Perez-Murano, Arantxa Fraile Rodríguez, and Xavier Batlle

Cite This: *ACS Nano* 2023, 17, 8123–8132 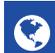 Read Online

ACCESS | 📊 Metrics & More | 📄 Article Recommendations | 🅢🅘 Supporting Information

**ABSTRACT:** Plasmonic lattice nanostructures are of technological interest because of their capacity to manipulate light below the diffraction limit. Here, we present a detailed study of dark and bright modes in the visible and near-infrared energy regime of an inverted plasmonic honeycomb lattice by a combination of Au⁺ focused ion beam lithography with nanometric resolution, optical and electron spectroscopy, and finite-difference time-domain simulations. The lattice consists of slits carved in a gold thin film, exhibiting hotspots and a set of bright and dark modes. We proposed that some of the dark modes detected by electron energy-loss spectroscopy are caused by antiferroelectric arrangements of the slit polarizations with two times the size of the hexagonal unit cell. The plasmonic resonances take place within the 0.5−2 eV energy range, indicating that they could be suitable for a synergistic coupling with excitons in two-dimensional transition metal dichalcogenides materials or for designing nanoscale sensing platforms based on near-field enhancement over a metallic surface.

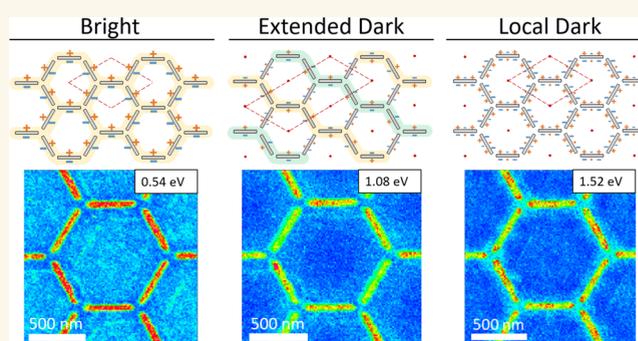

L ocalized surface plasmons (LSP) can be excited at the interface between a metallic nanostructure and a dielectric medium by coupling with an external electromagnetic wave under the appropriate conditions.[1] LSP excitations create subwavelength confinement of the light in the vicinity of the nanostructures, as well as an enhanced intensity of such a near-field distribution.[2−4] Consequently, they have extensively been used in a wide variety of applications involving electromagnetic radiation in the infra-red-visible-ultraviolet range, such as nanoantennas,[5] high sensitivity spectroscopies,[6,7] biomedical applications,[8−10] non-linear harmonic generation,[11] ultrafast phase modulation,[12] and sensors[13] among others.

An additional level of complexity arises when metallic nanostructures are organized in a periodic array. In this case, the radiative coupling between the LSP and the diffracted waves in the plane of the array enables the appearance of coherent excitations of the scatterers in the lattice, known as surface lattice resonances (SLR). Introduced by Carron and co-workers[14] and found experimentally by Hicks and co-workers,[15] the excitation of these resonances occurs near the frequency at which the diffracted wave is radiating in the plane of the array, that is, at the Rayleigh anomaly. Experimental realizations of lattices of plasmonic nanostructures vary considerably, both in design and functionality.[16−18] Although the LSP of neighboring nanoparticles can also be coupled, the corresponding excitations are severely affected by radiative damping. SLR, in contrast, often exhibit highly tunable, intense and narrow resonances.[19]

In this work, we focus on geometrically frustrated honey-comb lattices, which have previously been studied for its unique plasmonic band structure and its similarities with the behavior of graphene and other 2D materials.[20−22] We investigate the plasmonic properties of an inverted honeycomb lattice (inverted as the nanostructure is carved in a continuous layer). The interest for inverted structures arises, among other applications, from further increasing the sensitivity of surface enhanced Raman spectroscopy (SERS) to ultrasmall amounts of a given analyte,[20,23] which is relevant for the detection of diseases in early stages[24] and traces of contaminants in wastewater.[25] According to Babinet's principle[21,22], the optical response of a plasmonic nanostructure must be equivalent to that of its inverted counterpart. Besides, the excitation of SLR is perfectly possible in these type of inverted systems.[20] This enables designing inverted structures with a very similar spectral behavior to their counterparts, in terms of the excitation energy and spectral features of the resonances - while corresponding to a very different near-field distribution around the structure.[22,26,27]

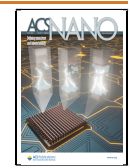

Here we show that the manufacture of a honeycomb array of slits fosters the formation of out-of-plane hotspots related to the SLR that present large enhancement factors of the electric field, even hundreds of nanometers away from the 2D array into the surrounding medium. Such an out-of-plane electric field enhancement is crucial for their integration in sensing and advanced spectroscopic architectures such as refractive index sensors.[28]

The implementation of any design pathway for enhanced plasmonic nanoarchitectures requires an understanding of the plasmonic near-field response. A possible route to obtain a clear picture of the plasmonic resonances supported by a particular structure is to use electron energy loss spectroscopy (EELS). This technique, probing the local photonic density of states,[29] allows the excitation and mapping of the local distribution of electromagnetic modes by measuring a spectrum of the energy loss resulting from its interactions with the sample, thereby becoming an important tool for near-field imaging of plasmonic optical excitations.[30] When performed in a monochromated aberration-corrected scanning transmission electron microscope (STEM), the achieved energy resolution of EELS can be of the order of 10 meV while preserving sub-Å spatial resolution.[31−34] A crucial advantage of this technique is the combination of the spectral features with their 2D projected intensity maps, allowing high sensitivity to subtle spatial modifications.[35] In the case of studying plasmonic responses with a STEM, an acquired EELS signal is closely related to the optical extinction spectra, and has proven to be a useful tool for studying localized and surface plasmon resonances of a variety of structures, including films, pillars, and holes of varying diameters, as well as slots and coaxial resonators,[36] metallic nanostructures,[37−39] nanocavities,[40,41] nanowires,[42] surface-plasmon modes in nanoparticles,[43,44] coupled nanoparticles,[45] molecular excitations,[46] or excitation of modes in 3D.[47] In addition to the experimental contributions, advances in EELS simulations and theoretical modeling have also been crucial for interpretation of the results.[35,48−50]

A key aspect regarding our work is the fact that, in addition to bright modes, EELS can also reveal optically dark modes,[35,36,51−54] that is, modes with a vanishing net dipole moment, thereby providing a full modal spectral map of a plasmonic system. Dark modes can store electromagnetic energy more efficiently than bright modes due to suppression of radiative losses. This results in narrower line widths and longer lifetimes than their radiative counterpart, making them ideal candidates for lossless nanoscale waveguides and subwavelength high-Q optical cavities,[55] as well as enhanced biological and chemical sensors or nanolasing applications.[56−59]

In this work, we focus on the excitation and mapping of the bright and dark plasmonic resonances associated with the SLR

and LSP modes in the optical range, which was achieved through the combination of state-of-the-art manufacturing of the samples and the outstanding spectral and spatial resolution of the STEM. A comprehensive spectral analysis of both the far-field and near-field measurements from Fourier-transform infrared spectroscopy (FTIR) and EELS, respectively, and the good agreement with the numerical electromagnetic simulations have enabled the identification of a variety of both bright and dark modes, the latter unable to be detected using light. Furthermore, dark modes that are be caused by arrangements of the slit polarizations that divide the system into two antiferroelectric sublattices have been found. These modes present a primitive unit cell twice that of the fabricated honeycomb lattice.

## RESULTS AND DISCUSSION

The studied sample consists in a 20 nm thick continuous layer of Au deposited on a thin (50 nm) $Si_3N_4$ membrane. The Au layer is patterned by carving slits through the entire thickness of the Au layer. The slits are approximately 400 nm long and 45 nm wide. They follow the arrangement of a honeycomb lattice with a pitch of p = 906 nm. This structure can be considered as a dielectric (air) planar honeycomb lattice embedded in a metallic (Au) layer (see Figure 1).

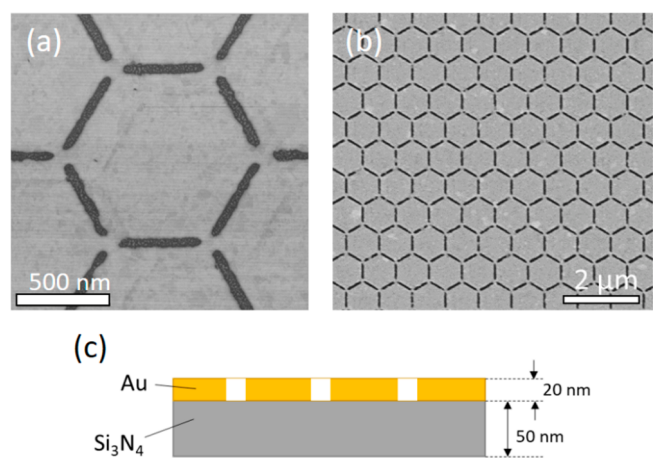

**Figure 1.** (a, b) Scanning electron microscopy images of the manufactured sample. (c) Schematic cross-sectional view of the studied sample.

The response of the structure presents a plethora of resonances in the visible and near-infrared (NIR) ranges. This is clear from the finite-difference time-domain (FDTD) simulations, FTIR measurements, and the data acquired during the EELS experiments. The corresponding spectral results are presented in Figure 2. Prior to an in-depth discussion, it is important to stress the differences between the three data sets shown in Figure 2.

FDTD simulations and FTIR measurements show the response of the system under the excitation of unpolarized light with normal incidence onto the array. However, EELS data rely on the energy-loss of a highly monochromatic beam of electrons that are inelastically scattered by the system. Therefore, FDTD simulation and FTIR data show only peaks associated with bright modes since electromagnetic radiation can only couple with modes exhibiting a net dipole moment.[55] In contrast, the scattering of electrons in an EELS experiment will excite all the available resonances in the structure,[50,60] including not only bright modes but also dark ones (those showing zero net dipole moment). This constitutes a key point in understanding the differences and similarities among the three curves shown in Figure 2. We also point out that the peaks shown by the FDTD simulated absorption do not exactly match the location of those on the FTIR and EELS curves because of the imperfections of the experimental sample due to manufacturing defects, such as slight variations in size or vertical profile of the slits.

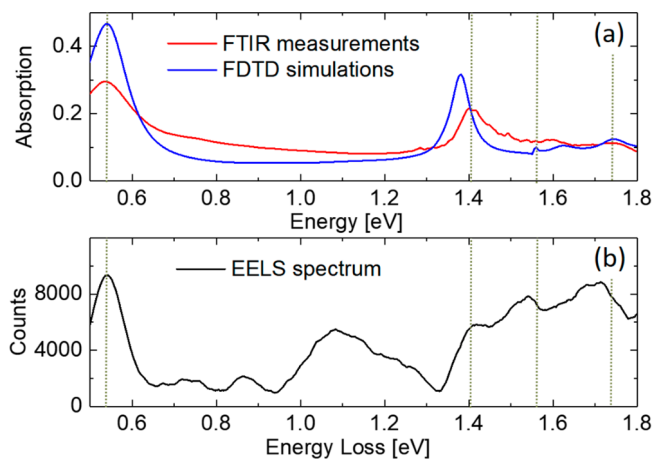

**Figure 2.** Absorption spectra obtained through (a) FDTD simulations (blue line) and FTIR measurements (red line). (b) EEL spectrum computed as the integral of the counts across the whole hexagon (black line). Peaks and anomalies only present in the EELS spectrum are associated with the excitation of dark modes by electron scattering. EELS and FTIR measurements were performed on the same sample. Vertical dashed lines indicate the energy of the bright modes that are discussed in the text.

The lowest energy mode of the honeycomb array of slits appears around 0.54 eV and corresponds to a relatively intense broad peak that is clearly visible in the three spectra shown in Figure 2. It is a dipolar mode caused by the distribution of opposite charges round the longitudinal facing edges of each slit, as it is disclosed by the FTDT simulations of charge distributions shown in Figure 3b. Moreover, it is a bright mode since the net dipole moment corresponding to the three slits emanating from every vertex (primitive unit element of the lattice) in Figure 3b is nonzero along the vertical direction parallel to the polarization axis of the exciting radiation in the FDTD simulations. Note that the charge distributions shown in Figure 3 were obtained from an FDTD simulation with linearly polarized light in the vertical direction. EELS measurements and the distribution of the intensity of the electric field for unpolarized light obtained from FDTD simulations support the occurrence of significant values of the electric field only in the hollow space inside the slits (Figure 3a and c) in accordance with the major dipolar nature of the mode in the directions perpendicular to the slits.

To get a deeper insight into the energy of the rest of the bright modes, the values of the integral of the EELS signal computed in several specific regions of the hexagon are shown in Figure 4 as a function of the energy loss within 1.2 and 1.8 eV, the energy range where most of the bright modes appear.

The next bright mode can be found at about 1.41 eV in the EELS and FTIR curves, and at 1.38 eV in the simulated absorption (see Figure 2). At 1.41 eV, there is also a clear peak in the light brown curve in Figure 4a (region III), as well as some anomalies in the curves for region I in Figure 4a. Based on the results of the EELS mapping and the FDTD simulations for unpolarized light, the near-field distribution of this resonance presents a diffuse hotspot around the center of each hexagon accompanied by a multipolar excitation of the slits (Figure 3d and f).

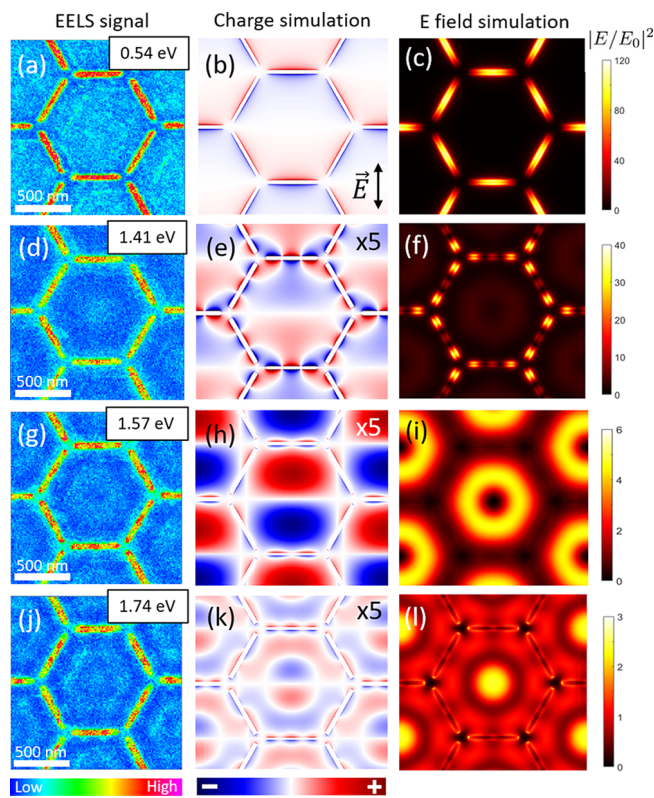

**Figure 3.** Left hand-side panels (a, d, g, and j) show colormaps of the EELS signal across the sample for the four bright modes indicated in Figure 2. The color scale bar represents the intensity of the EELS signal. Middle panels (b, e, h, and k) depict the calculated charge distribution on the surface of the array for the same bright modes from the FDTD simulations under the illumination of linearly polarized light along the vertical axis. Blue and red colors represent negative and positive net charge densities, respectively. The charge intensity has been multiplied by a factor, indicated by the number on the top-right side of the panels, for an easier comparison between resonances. Right hand-side panels (c, f, i, and l) show simulations of the corresponding electric field intensity under the illumination of unpolarized light. These colormaps were obtained 10 nm above the structure.

The charge distribution shown in Figure 3e, which was obtained from an FDTD simulation with linearly polarized light in the vertical direction, also indicates a multipolar excitation of the slits. Each slit contains three parallel dipoles of alternating sign. So, as in the case of the previous excitation at 0.54 eV, every three converging slits at each vertex has a net dipole moment in the vertical direction. At the same time, the inferior and superior halves of the hexagons exhibit diffuse charges of opposite sign that contribute to the total dipole moment of the structure in the vertical direction (Figure 3e) and cause the central diffuse hotspot found under unpolarized radiation (see Figure 3d and f).

The mode at 1.57 eV corresponds to the coherent excitation of the whole lattice in a SLR. The small peak in the absorption spectrum obtained by FDTD simulations around this energy (blue curve in Figure 2a) is associated with the diffraction condition of the lattice at the interface between the metal and the air. In fact, a photon of 1.57 eV in air has a wavelength of 790 nm, which is very close to $p \cos 30 = 785$ nm, the first diffraction condition of the lattice for normal incidence, being p the pitch of the lattice. Interestingly, the imperfections of the experimental sample do not affect its pitch (see Figure 1a and b). This means that the SLR of the lattice takes place at the same energy as the FDTD simulations and experimental results. In this case, most of the excited charge is found to spread inside the hexagons, relatively far from the slits forming the array (see Figure 3h for FDTD results with linearly polarized light). The two opposite sign charge distributions in the upper and lower halves of each hexagon cause a large electric field with an out-of-plane component that extends hundreds of nanometers from the structure[28] (see Figure S1 in the Supporting Information). For unpolarized radiation with normal incidence, the charge distribution will oscillate from the center of the hexagon to the outer part following a kind of breathing mode and giving rise to the EELS signal and electric-field distribution depicted in Figure 3g and i. Moreover, the $xy$- plane crosscut for the $z$-component of the electric field shown in Figure S1 in the Supporting Information is also characteristic of a breathing mode.

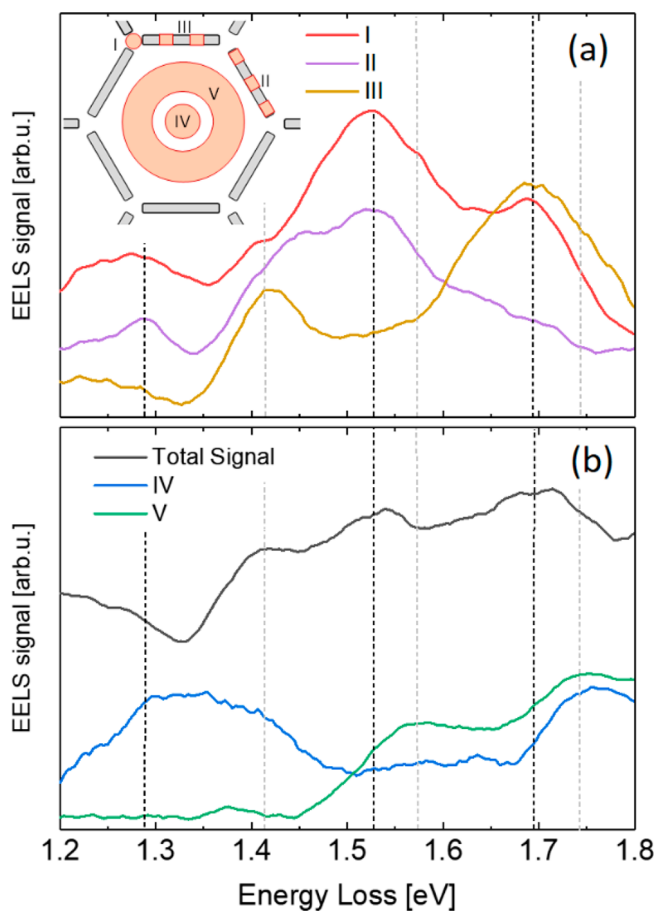

**Figure 4.** EEL spectra computed as the integral of the counts across the regions of the hexagon colored in orange in the inset of panel (a). (a) Red, purple, and light brown solid curves correspond to the integrals of the EELS signal over regions I, II, and III, respectively. (b) Blue and green solid curves correspond to the integrals of the EELS signal over regions IV and V, respectively. The total signal across the whole hexagon is also shown as the black solid line in panel (b) for the sake of comparison. The vertical dashed lines indicate the energy of the bright (gray lines) and dark (black lines) modes discussed in the text.

Besides, for this mode, the intensity of the electric field for unpolarized light shows a central hotspot with a ring shape, revolving around a minimum of intensity at the center of the hexagon (see Figure 3i). It is worth noting that a central structure like that of Figure 3i is clearly distinguishable in the EELS colormap (Figure 3g). This qualitative interpretation of the EELS colormap is also quantitatively confirmed by the local maximum of the green curve in Figure 4b corresponding to the integral of the EELS signal in region V (central ring). It should also be noted that, in relation to this mode, a small shoulder is shown in the red curve in Figure 4a (region I corresponding to the vertex).

Finally, at an energy of 1.74 eV we find an excitation in the EELS experiment (Figure 3j) with a maximum in the center of the hexagons surrounded by an approximately hexagonal hotspot. This central structure may appear due to an averaging along the directions perpendicular to the edges of the hexagon of charge distributions such that in Figure 3k for linearly polarized light in the vertical direction. The existence of this central structure can also be confirmed by the maxima in the green and blue curves in Figure 4b for the EELS signal integrated in regions V (central ring) and IV (central hotspot), respectively. In addition, the EELS map also suggests a multipolar excitation of the slits in agreement with the FDTD simulation for the electric field distribution in Figure 3l.

The four modes discussed so far are bright, since they display a net dipole moment per primitive unit cell of the hexagonal lattice. In this respect, they can be considered ferroelectric modes of the system, since the dipole moment per primitive cell forms a ferroelectric lattice with the same symmetry as that of the hexagonal array of slits.

In addition, dark modes are also excited in the structure. In this work, we present the detection of two types of dark modes. Depending on whether the unit cell of the charge distribution fits into a single primitive cell of the honeycomb lattice, or it extends outside of it, we will denote these modes as local or extended, respectively. A first approach to understanding the local charge distributions excited in a dark mode is to consider the behavior of three slits converging on the same vertex (Figure S2 in the Supporting Information) that will act as the unit element of the lattice. Bearing in mind that we are dealing with an inverted lattice, most of the excitation of this unit element will take place through currents flowing around the central vertex, giving rise to a central out-of-plane magnetic moment.[61] Therefore, by placing a perpendicular magnetic dipole source above the vertex shared by these three slits, several dark modes of increasing multipolar order can be simulated as the energy increases. The lowest energy dark mode (Figure S2a in the Supporting Information) shows a dipolar excitation across the slits, akin to the bright mode at 0.54 eV but, in this case, all the slits are equivalent due to the 3-fold symmetry of the element, and the net dipole moment vanishes. The other local dark modes shown by the simulations as the energy increases follow the same overall behavior but with multipolar excitation of the slits (see Figure S2b,c in the Supporting Information).

Homologous dark modes can be found in the inverted honeycomb lattice by exciting the system with an array of in-phase magnetic dipoles situated on those vertices of the hexagons that coincide with the Bravais lattice (see Methods section). The lowest energy dark mode found by simulation is shown in Figure S3 in the Supporting Information. However, due to the similarity between the near-field distributions of this mode and that of the broad bright one at 0.54 eV and the spectral proximity of both, this excitation cannot be distinguished in the EELS spectrum in Figure 2. The rest of the local dark modes are presented in Figure 5, in order of increasing energy. Figure 5b shows the average of the profile of the EELS signal at 1.28 eV computed as the integral across the slits, where the purple curve in Figure 4a (EELS signal from the ends and center of the slits) exhibits a local maximum. This profile supports the fact that most of the excitation occurs at both ends of the slits following a quadrupolar polarization, and is in good agreement with the simulated charge and near-field distributions in Figure 5c,d. We point out that any quadrupolar arrangement of the charge distribution around the slits nonhaving 3-fold symmetry gives rise also to dark modes of similar energies.

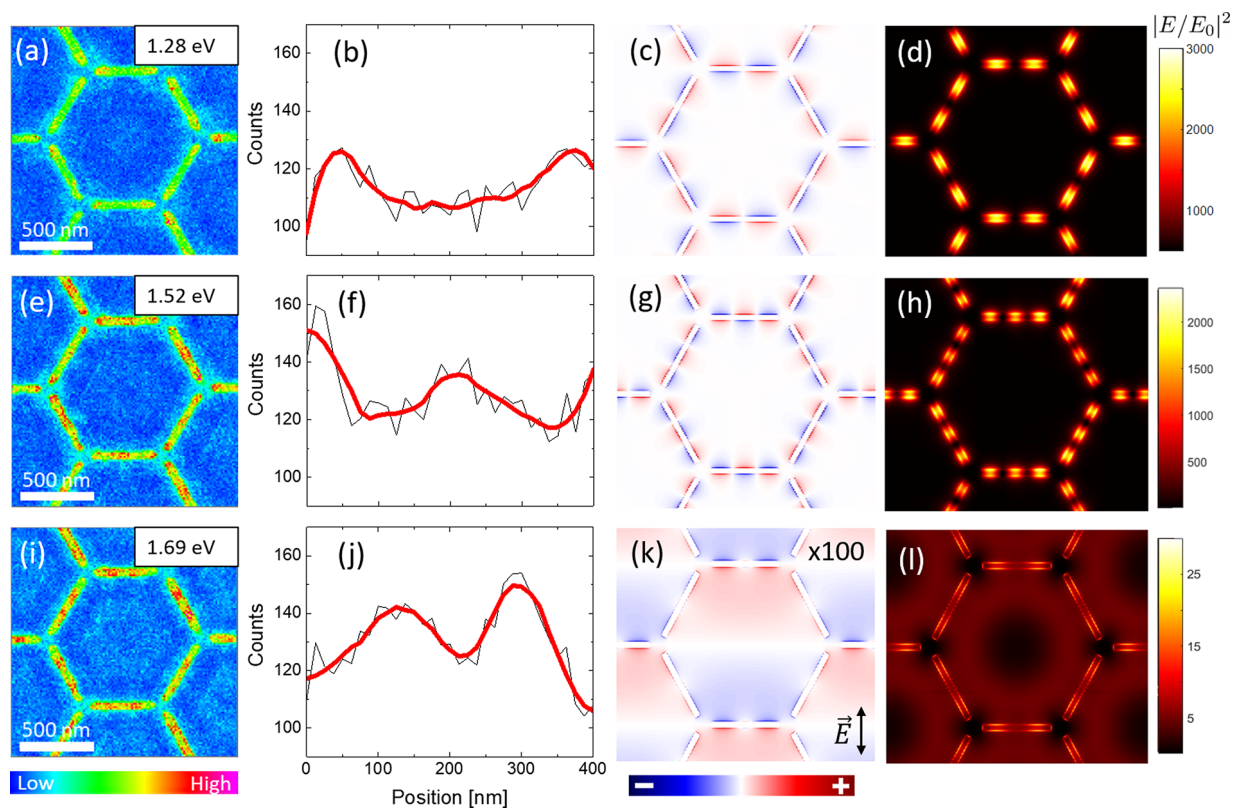

**Figure 5.** Left hand-side panels (a, e, and i) show colormaps of the EELS signal across the sample for the three dark modes indicated in Figure 4a. The color scale bar represents the intensity of the EELS signal. Panels (b, f, and j) depict the average of the profile of the EELS signal along the slits computed integrating the counts across the slits. Panels (c, g, and k) show the simulated charge distributions of these three dark modes. Right hand-side panels (d, h, and l) show the near field distribution for each energy computed 10 nm above the structure. Note that for the modes arising at 1.28 and 1.52 eV an array of magnetic dipoles has been used to excite the system, whereas for the mode at 1.69 eV the system has been excited using a combination of an array of magnetic dipoles and a plane wave with normal incidence. Panel (k) shows the charge distribution for the vertical polarization of the plane wave while panel l is the average between two orthogonal polarizations. These colormaps were obtained 10 nm above the structure.

At 1.52 eV there are local maxima in the red (EELS signal at the vertices of the hexagonal lattice) and purple (EELS signal from the ends and center of the slits) curves in Figure 4a that are associated with a dark mode with the sextupole polarization of the slits. Accordingly, the respective EELS profile along the slits (see Figure 5f) shows three approximately equidistant poles. The results of the simulations using magnetic dipole sources shown in Figure 5g,h support this interpretation.

Finally, at about 1.69 eV there are maxima in the red (EELS signal at the vertices of the hexagonal lattice) and light brown (EELS signal from regions of the slits that exclude their ends and center) curves in Figure 4a that coincide with a broad maximum in the curve for the total signal (black solid line in Figure 4b). In accordance with the results of the simulations with magnetic dipoles, this should be a dark mode with octupole polarization of the slits and relatively low intensity. However, only two poles close to the center of the slits are shown by both the EELS color map in Figure 5i and the profile of the EELS signal along the slits in Figure 5j. In addition, EELS color map in Figure 5i shows a distinctive ring around the center of the hexagon. Such an electric field pattern does not arise from the sole excitation of charge across the slits, but from a charge arrangement in the continuous gold layer found in the hexagons formed by the honeycomb lattice of slits. Therefore, this mode may result from the hybridization of the octupole dark mode and a bright mode like that at 1.74 eV. To simulate this mode, we have used the simultaneous excitation of the system by an array of in-phase magnetic dipoles like in the previous cases and a plane wave with normal incidence, aiming at exciting both dark and bright modes. The charge distribution depicted in Figure 5k, which was obtained for the polarization axis of the plane wave along the vertical direction, shows a kind of quadrupolar excitation close to the center of the horizontal slits (those perpendicular to the polarization axis of the plane wave) that is compatible with the EELS profile shown in Figure 5j. Moreover, the near-field distribution computed for an unpolarized plane wave in Figure 5l shows overall features in qualitative agreement with EELS color map in Figure 5i, including the ring around the center of the hexagon and the two central poles in the slits.

In addition to the local dark modes discussed so far, there are several other ways to get an extended dark mode in an ordered lattice. For instance, the system can be divided into two sublattices of opposite net dipole moment per primitive unit cell, so that the net dipole moment cancels out in pairs of two primitive cells. This implies that there is a net dipole moment associated with each threesome of slits converging in a vertex coincident with the Bravais lattice (unit element of the lattice) that cancels out with the opposite net dipole moment of a neighboring threesome of slits. Therefore, the excitation of the slits in each of these threesomes must be nonequivalent under the 3-fold symmetry to have a net dipole moment, like in the bright mode.

Examples of this kind of dark modes can be simulated by FDTD using a tailored source of radiation that excites neighboring primitive unit cells in opposition of phase, in such a way that they form two antiparallel ferroelectric sublattices. An array of magnetic dipoles in phase opposition and placed above the vertices coincident with the Bravais lattice was the excitation source used for this purpose (see more details in the Methods section). Thus, Figure 6a,b shows an example of these extended dark modes at 0.68 eV where the charge across the slits alternates along the zigzag chains, and the remaining slits, rotated 60 degrees from the horizontal, are less charged and show a kind of quadrupolar excitation. This results in two antiferroelectric sublattices whose net dipole moments (indicated by arrows in Figure 6b) per primitive cell of the honeycomb lattice point in the direction of the zigzag chains and are arranged antiparallel to each other, so they cancel out in pairs of primitive cells. Note that the unit cell of the antiferroelectric mode is indicated by the dashed black line in Figure 6b and is two times bigger than that of the honeycomb lattice. It is worth noting that for this kind of extended dark modes, one of the slits of each threesome shows a different excitation (different multipolarity and/or intensity) than the other two, as expected. This is a distinctive feature of these extended dark modes that can be checked in the results of the simulations for the near-field distribution at other energies that are shown in Figure S4 in the Supporting Information. Consequently, there exists an almost continuous family of extended dark modes with increasing energy, where the differences among them come from the multipolarities of the slits in each threesome, and/or their relative intensities of excitation.

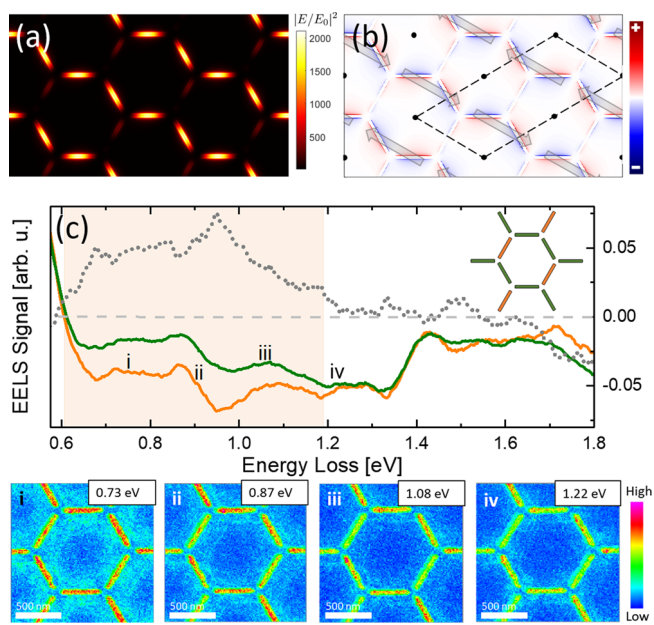

Figure 6. (a, b) Simulated electric field and charge distribution, respectively, corresponding to the antiferroelectric mode at 0.68 eV. The arrows in panel (b) indicate the total electrical dipole moments for threesomes of slits converging at points of the Bravais lattice. The dashed lines show a primitive unit cell of the antiferroelectric mode. (c) Average integral of the EELS signal computed separately over the two subsets of slits (the green and orange curves correspond to the green and orange subsets of slits in the inset showing a unit cell of the lattice). EELS color maps labeled with Roman numerals show some representative examples of the extended dark modes taking place in the energy range within 0.6 and 1.22 eV.

This kind of extended dark mode can be excited by EELS in the experimental sample as discussed below. Previous observations of local dark modes by EELS, among other techniques, have been reported.[36,45,62] Nevertheless, the finding of extended (antiferroelectric) dark modes is also reported in this work.

Between 0.6 and 1.4 eV, there are no bright modes in the spectral response of the system, so all the peaks and anomalies shown by the EELS signal in this range in Figure 2b could be assigned to dark modes. To find out the nature of these dark modes, we represent in Figure 6c the average integral of the EELS signal as a function of energy computed separately over two subsets of slits (those forming zigzag chains and the remaining ones, colored in green and orange in inset of Figure 6c, respectively). Interestingly, the EELS signals for the green and orange slits coincide for the bright mode at 0.54 eV, as expected. But as the energy increases and becomes larger than 0.6 eV, the two curves in Figure 6c start to separate and stay that way until around 1.2 eV, from where they coincide again within the experimental error. In this energy range (light brown region in Figure 6c) the slits forming the zigzag chains (shown in green in the inset of Figure 6c) are excited more strongly than the remaining isolated slits (shown in orange in the inset of Figure 6c), indicating that the system cannot be excited through local dark modes, since, if that were the case, the intensity of excitation of the slits in each threesome would be equal. Consequently, there should be a net dipole moment associated with each threesome of slits converging in a vertex since the excitations of them are not equivalent under the 3-fold symmetry, and the system may be excited through a kind of extended dark mode, as described above. However, it is worth noting that, in an EELS experiment with an ideal sample, all the mode variants equivalent by symmetry (three in a honeycomb lattice) are simultaneously excited with the same probability, in such a way that the resulting state exhibits always the same average intensity and multipolarity for all the slits, even for an extended dark mode. In our case, the imperfections in the fabrication of the sample (especially in the oblique slits) enable a distinct excitation of the three variants of the extended mode, yielding an average state where the signature of an extended dark mode can be inferred by the different excitation intensity of the three slits of each threesome.

The EELS color maps labeled with Roman numerals i, ii, iii, and iv in Figure 6c show some representative examples of the excitation of the system corresponding to three local maxima of the EELS signal in this energy range and the limit of the energy range where extended dark modes can be excited, respectively. The EELS color maps i and ii in Figure 6c and the EELS profiles along the slits in Figure S4,a,b in the Supporting Information indicate a dipolar excitation of the slits but with stronger intensity for the slits forming zigzag chains. The results of the simulations for the electric field and the average profile along the slits obtained with the array of magnetic dipoles in phase opposition (Figure S4e,f,i,j in the Supporting Information) are in qualitative agreement, since the average of the excitation of the three slits in a threesome has major dipolar character. On the contrary, the EELS color maps iii and iv in Figure 6 and the EELS profiles along the slits in Figures S4,c,d in the Supporting Information indicate a quadrupole excitation of the slits. The corresponding results of the simulations for the electric field and the average profiles along the slits (Figure S4g,h,k,l in the Supporting Information) point toward the quadrupole character of the average excitation of the slits. Therefore, as the energy increases from 0.6 eV, the average polarity of the extended dark modes progressively becomes of higher order, until the local dark mode corresponding to quadrupole excitation with the same intensity of all the slits in a threesome is reached at 1.28 eV.

At energies higher than about 1.7 eV, the EELS signals for the green and orange slits in the inset of Figure 6c start to significantly separate again, indicating the onset of a second range of energies where antiferroelectric dark modes may be excited, but with the intensity of the excitation of the two subsets of slits swapped with respect to those within 0.6 and 1.2 eV.

## CONCLUSIONS

This work constitutes a comprehensive study of the plasmonic properties of an inverted honeycomb lattice of slits. The patterning quality of the samples, together with the spectral and spatial resolution of the EELS measurements has led to the direct observation and mapping of bright and dark plasmonic modes. A detailed description of the charge and near-field distributions in the structure has been given by virtue of the good agreement between the EELS measurements, the optical measurements, and simulations. Some of the dark modes found are caused by antiferroelectric arrangements of the slit polarizations, giving rise to charge arrangements with a unit cell two times larger than that of the original honeycomb lattice. Additionally, plasmonic modes exhibiting hotspots far from the discontinuities of the metallic layer are found, ranging from 1.3 to 1.8 eV approximately.

The behavior of the inverted honeycomb plasmonic lattice is relevant not only from a fundamental point of view. As shown in a previous work,[24] the appearance of hotspots far from the slits that form the lattice is highly correlated with a strong out-of-plane electric field ranging hundreds of nanometers away from the lattice. This electric field could foster a strong coupling of the plasmonic lattice with other materials allocated on top of the structure. Moreover, the inverted nature of this lattice, thanks to its large mostly planar surface, presents this system as a platform for exploiting new synergies with 2D materials. The exciton energies for 2D $WSe_2$ and $MoS_2$ on an Au substrate, 1.75 and 1.9 eV, respectively,[63] could be targeted by easily tuning manufacturing parameters such as the pitch of the lattice, thus changing the spectral position of the plasmonic resonances.

Other applications include transport measurements in 2D materials in which a local gating could be done by means of the hotspot in the nanostructure and the introduction of localized defects in the lattice.

## METHODS

**Manufacture Process.** A thin Au film was deposited on a 50 nm thick $Si_3N_4$ membrane using an ultrahigh vacuum (UHV) electron beam evaporator. The $Si_3N_4$ membrane is $500 \times 500 \ \mu m^2$ wide. These types of structures are normally used as substrate for Transmission Electron Microscopy (TEM). The square membrane windows are centered on 200 $\mu m$ thickness silicon frames.[64]

The measured Au film thickness was 19.4 nm with a root-mean-square roughness of 0.3 nm as measured by atomic force microscopy (AFM). The Au film was then patterned using a Raith Velion focused ion beam (FIB) equipped with an Au−Ge−Si liquid metal alloy ion source (LMAIS). A 35 keV Au+ beam, ∼5 pA ion current intensity and ∼15 nm diameter, was used for patterning the hexagonal array of slits. Au+ beam was chosen to avoid lateral contaminations along the milled lines that would occur with other commonly used ion species in FIB like Ga+. The $350 \times 350 \ \mu m^2$ wide array was milled using $200 \times 200 \ \mu m^2$ writing fields and applying a 3000 pC/cm linear dose of Au+ (single loop passage, 10 nm step between ion spots, and dwell time of 0.52 ms). This allowed for the patterning of the designed structures through the Au film while keeping the line width in plane of ∼45 nm.

**FTIR Measurements.** The optical characterization was carried out using a Vertex 70 Fourier transform infrared (FTIR) spectrophotometer attached to an optical microscope (Bruker Hyperion). Experiments were performed in the reflection and transmission configuration with 2× and 4× objectives, respectively, under unpolarized light illumination. A shutter was used to select the signal coming from the nanostructured area. The measured signal of the plain Au of a non-nanostructured area of the sample was used as a background for the reflection measurements, whereas the transmission of the light in air was used for calibrating the transmission measurements.

**EELS Measurements.** EEL spectra were collected using a Nion aberration-corrected high energy resolution monochromated EELS−STEM (Nion HERMES) operating at a 60 kV accelerating voltage, using a convergence semiangle of 30 mrad, a collection semiangle of 20 mrad, and a beam current of ∼10 pA.[52−54] The resulting energy resolution of the spectra, measured by the full-width half-maximum (fwhm) of the zero-loss peak was 60 meV.

**FDTD Simulations.** The simulations were performed using the finite-difference time-domain (FDTD) method, implemented in the solver provided by Lumerical.[65] The computation procedure starts by setting a 3D FDTD simulation with perfectly matched layers (PMLs) in the $z$-direction, perpendicular to the plane of the lattice and periodic boundary conditions in the other two directions, the $x−y$ plane. Given the initial conditions, the software solves Maxwell's equations to determine the Fourier components of the electric and magnetic fields by using discrete time and spatial steps. The method allows for a direct observation of the physical phenomena taking place without imposing any further assumption on the behavior of the system. We obtained the transmission and reflection spectra by placing two monitors that compute the total power that flows through a surface. In addition, other monitors were placed on the surface of the sample and 10 nm above, from which we computed the charge and the electric field distributions, respectively.

An override mesh region was defined along the structure to ensure a correct level of detail in the description of the lattice. The cell size used was $3.816 \times 3.824 \times 1 \ nm^3$ in the $x$-, $y$- and $z$-directions, respectively. The dielectric functions for the materials used in these simulations were obtained by fitting analytical functions to the data from refs [66] and [67] for $Si_3N_4$ and Au, respectively.

In these simulations, four different types of excitation sources were used, depending on the targeted modes. Plane waves were the source of excitation for all the bright modes. The plane waves were injected at 500 nm from the structure following normal incidence and setting periodic boundary conditions in the $x-y$ plane for a unit cell of the system like that of the snapshots in Figure 3. For the simulations showing local dark modes, an array of in-phase magnetic dipoles, perpendicular to the system plane and placed 50 nm above the vertices of the hexagons coincident with the Bravais lattice, was the excitation procedure. Bloch boundary conditions in the x-y directions were set for the same unit cell than the one for the bright modes. The hybrid mode at 1.69 eV was simulated by the simultaneous excitation of a plane wave with normal incidence and the array of in-phase magnetic dipoles. Finally, for the antiferroelectric dark modes, an array of magnetic dipoles in phase opposition were placed 50 nm above the vertices coincident with the Bravais lattice (see Figure S5 in the Supporting Information). The simulation unit cell, as shown in Figure S5 in the Supporting Information, was larger than in the previous cases to enable the excitation of extended modes. Bloch boundary condition were used in the x-y directions.

## ASSOCIATED CONTENT

### ⓈSupporting Information

The Supporting Information is available free of charge at https://pubs.acs.org/doi/10.1021/acsnano.2c11016.

Transversal electric field distribution for the SLR at 1.57 eV; simulated electric field and charge distributions for a threesome of slits; simulated electric field and charge distributions for the simplest local dark mode of the inverted honeycomb lattice; profiles of the EELS signal and the simulated electric field along the slits for the antiferroelectric dark modes; array of the magnetic dipoles over the structure used to simulate antiferroelectric dark modes (PDF)

## AUTHOR INFORMATION

### Corresponding Author
**Javier Rodríguez-Álvarez** − *Departament de Física de la Matèria Condensada, Universitat de Barcelona, Barcelona 08028, Spain; Institut de Nanociència i Nanotecnologia (IN2UB), Universitat de Barcelona, Barcelona 08028, Spain;* ⓞ orcid.org/0000-0001-5822-4013; Email: javier.rodriguez@ub.edu

### Authors
**Amílcar Labarta** − *Departament de Física de la Matèria Condensada, Universitat de Barcelona, Barcelona 08028, Spain; Institut de Nanociència i Nanotecnologia (IN2UB), Universitat de Barcelona, Barcelona 08028, Spain;* ⓞ orcid.org/0000-0003-0904-4678

**Juan Carlos Idrobo** − *Materials Science and Engineering Department, University of Washington, Seattle, Washington 98195, United States*

**Rossana Dell'Anna** − *Sensors & Devices Center, FBK - Bruno Kessler Foundation, Povo, TN 38123, Italy;* ⓞ orcid.org/0000-0001-7147-6127

**Alessandro Cian** − *Sensors & Devices Center, FBK - Bruno Kessler Foundation, Povo, TN 38123, Italy*

**Damiano Giubertoni** − *Sensors & Devices Center, FBK - Bruno Kessler Foundation, Povo, TN 38123, Italy;* ⓞ orcid.org/0000-0001-8197-8729

**Xavier Borrisé** − *Catalan Institute of Nanoscience and Nanotechnology (ICN2), CSIC and BIST, Campus UAB, Bellaterra, Barcelona 08193, Spain*

**Albert Guerrero** − *Institut de Microelectrònica de Barcelona (IMB-CNM, CSIC), Bellaterra 08193, Spain*

**Francesc Perez-Murano** − *Institut de Microelectrònica de Barcelona (IMB-CNM, CSIC), Bellaterra 08193, Spain*

**Arantxa Fraile Rodríguez** − *Departament de Física de la Matèria Condensada, Universitat de Barcelona, Barcelona 08028, Spain; Institut de Nanociència i Nanotecnologia (IN2UB), Universitat de Barcelona, Barcelona 08028, Spain;* ⓞ orcid.org/0000-0003-2722-0882

**Xavier Batlle** − *Departament de Física de la Matèria Condensada, Universitat de Barcelona, Barcelona 08028, Spain; Institut de Nanociència i Nanotecnologia (IN2UB), Universitat de Barcelona, Barcelona 08028, Spain;* ⓞ orcid.org/0000-0001-7897-2692

Complete contact information is available at:
https://pubs.acs.org/10.1021/acsnano.2c11016

### Author Contributions
The manuscript was written through contributions of all authors. All authors have given approval to the final version of the manuscript.

### Notes
The authors declare no competing financial interest.

## ACKNOWLEDGMENTS

The authors want to thank Pau Molet and Agustín Mihi, Institut de Ciència de Materials de Barcelona (ICMAB-CSIC), for the FTIR measurements, and Paolo Mattevi of FBK for the UHV evaporation of Au on the $Si_3N_4$ membranes. This work was supported by Spanish ICTS Network MICRONANO-FABS. The authors from Universitat de Barcelona thank the funding from Spanish MICIIN, grant numbers PGC2018-097789−B-I00 and PID2021-127397NB-I00, and the European Union FEDER funds. The EELS measurements were supported by the Center for Nanophase Materials Sciences (CNMS), which is a U.S. Department of Energy, Office of Science User Facility, and using instrumentation within ORNL's Materials Characterization Core provided by UT-Batelle, LLC, under Contract No. DE-AC05-00OR22725 with the U.S. Department of Energy, and sponsored by the Laboratory Directed Research and Development Program of Oak Ridge National Laboratory, managed by UT-Battelle, LLC, for the U.S. Department of Energy.

## REFERENCES

(1) Maier, S. A. *Plasmonics fundamentals an applications*, 1st ed.; Springer: New York, 2007.

(2) Marinica, D. C.; Kazansky, A. K.; Nordlander, P.; Aizpurua, J.; Borisov, A. G. Quantum plasmonics: Nonlinear effects in the field enhancement of a plasmonic nanoparticle dimer. *Nano Lett.* **2012**, *12*, 1333−1339.

(3) Lee, B.; Lee, I. M.; Kim, S.; Oh, D. H.; Hesselink, L. Review on subwavelength confinement of light with plasmonics. *J. Mod. Opt.* **2010**, *57*, 1479−1497.

(4) Khurgin, J.; Tsai, W. Y.; Tsai, D. P.; Sun, G. Landau Damping and Limit to Field Confinement and Enhancement in Plasmonic Dimers. *ACS Photonics* **2017**, *4*, 2871−2880.

(5) Novotny, L.; Van Hulst, N. Antennas for light. *Nat. Photonics* **2011**, *5*, 83−90.

(6) Maccaferri, N.; Barbillon, G.; Koya, A. N.; Lu, G.; Acuna, G. P.; Garoli, D. Recent advances in plasmonic nanocavities for single-molecule spectroscopy. *Nanoscale Adv.* 2021, 3, 633−642.

(7) Cho, W. J.; Kim, Y.; Kim, J. K. Ultrahigh-Density Array of Silver Nanoclusters for SERS Substrate with High Sensitivity and Excellent Reproducibility. *ACS Nano* 2012, 6, 249−255.

(8) Vo-dinh, T.; Wang, H.; Scaffidi, J. Plasmonic nanoprobes for SERS biosensing and bioimaging. *J. Biophotonics* 2010, 3, 89−102.

(9) Bauch, M.; Toma, K.; Toma, M.; Zhang, Q.; Dostalek, J. Plasmon-Enhanced Fluorescence Biosensors : a Review. *Plasmonics* 2014, 9, 781−799.

(10) Sharifi, M.; Attar, F.; Saboury, A. A.; Akhtari, K.; Hooshmand, N.; Hasan, A.; El-Sayed, M. A.; Falahati, M. Plasmonic gold nanoparticles: Optical manipulation, imaging, drug delivery and therapy. *J. Controlled Release* 2019, 311−312, 170−189.

(11) Ahmadivand, A.; Gerislioglu, B. Deep- and vacuum-ultraviolet metaphotonic light sources. *Mater. Today* 2021, 51, 208−221.

(12) Smolyaninov, A.; El Amili, A.; Vallini, F.; Pappert, S.; Fainman, Y. Programmable plasmonic phase modulation of free-space wavefronts at gigahertz rates. *Nat. Photonics* 2019, 13, 431−435.

(13) Ahmadivand, A.; Gerislioglu, B. Photonic and Plasmonic Metasensors. *Laser Photonics Rev.* 2022, 16, 2100328.

(14) Carron, K. T.; Lehmann, H. W.; Fluhr, W.; Meier, M.; Wokaun, A. Resonances of two-dimensional particle gratings in surface-enhanced Raman scattering. *J. Opt. Soc. Am. B* 1986, 3, 430.

(15) Hicks, E. M.; Zou, S.; Schatz, G. C.; Spears, K. G.; Van Duyne, R. P.; Gunnarsson, L.; Rindzevicius, T.; Kasemo, B.; Käll, M. Controlling plasmon line shapes through diffractive coupling in linear arrays of cylindrical nanoparticles fabricated by electron beam lithography. *Nano Lett.* 2005, 5, 1065−1070.

(16) Rodriguez, S. R. K.; Abass, A.; Maes, B.; Janssen, O. T. A.; Vecchi, G.; Gómez Rivas, J. Coupling Bright and Dark Plasmonic Lattice Resonances. *Phys. Rev. X* 2011, 1, 1−7.

(17) Humphrey, A. D.; Barnes, W. L. Plasmonic surface lattice resonances on arrays of different lattice symmetry. *Phys. Rev. B - Condens. Matter Mater. Phys.* 2014, 90, 1−8.

(18) Conde-Rubio, A.; Fraile Rodríguez, A.; Espinha, A.; Mihi, A.; Pérez-Murano, F.; Batlle, X.; Labarta, A. Geometric frustration in ordered lattices of plasmonic nanoelements. *Sci. Rep.* 2019, 9, 1−10.

(19) Vecchi, G.; Giannini, V.; Gómez Rivas, J. Surface modes in plasmonic crystals induced by diffractive coupling of nanoantennas. *Phys. Rev. B - Condens. Matter Mater. Phys.* 2009, 80, 1−4.

(20) Huck, C.; Vogt, J.; Sendner, M.; Hengstler, D.; Neubrech, F.; Pucci, A. Plasmonic Enhancement of Infrared Vibrational Signals: Nanoslits versus Nanorods. *ACS Photonics* 2015, 2, 1489−1497.

(21) Hentschel, M.; Weiss, T.; Bagheri, S.; Giessen, H. Babinet to the half: Coupling of solid and inverse plasmonic structures. *Nano Lett.* 2013, 13, 4428−4433.

(22) Hrtoň, M.; Konečná, A.; Horák, M.; Šikola, T.; Křápek, V. Plasmonic Antennas with Electric, Magnetic, and Electromagnetic Hot Spots Based on Babinet's Principle. *Phys. Rev. Appl.* 2020, 13, 054045.

(23) Cetin, A. E.; Turkmen, M.; Aksu, S.; Etezadi, D.; Altug, H. Multi-resonant compact nanoaperture with accessible large nearfields. *Appl. Phys. B: Laser Opt.* 2015, 118, 29−38.

(24) Moore, T.; Moody, A.; Payne, T.; Sarabia, G.; Daniel, A.; Sharma, B. In Vitro and In Vivo SERS Biosensing for Disease Diagnosis. *Biosensors* 2018, 8, 46.

(25) Wei, H.; Hossein Abtahi, S. M.; Vikesland, P. J. Environmental Science Plasmonic colorimetric and SERS sensors for environmental analysis. *Environ. Sci. Nano* 2015, 2, 120−135.

(26) Falcone, F.; Lopetegi, T.; Laso, M. A. G.; Baena, J. D.; Bonache, J.; Beruete, M.; Marques, R.; Martín, F.; Sorolla, M. Babinet principle applied to the design of metasurfaces and metamaterials. *Phys. Rev. Lett.* 2004, 93, 2−5.

(27) Zentgraf, T.; Meyrath, T. P.; Seidel, A.; Kaiser, S.; Giessen, H.; Rockstuhl, C.; Lederer, F. Babinet's principle for optical frequency metamaterials and nanoantennas. *Phys. Rev. B - Condens. Matter Mater. Phys.* 2007, 76, 4−7.

(28) Rodríguez-Álvarez, J.; Gnoatto, L.; Martínez-Castells, M.; Guerrero, A.; Borrisé, X.; Fraile Rodríguez, A.; Batlle, X.; Labarta, A. An inverted honeycomb plasmonic lattice as an efficient refractive index sensor. *Nanomaterials* 2021, 11, 1217.

(29) García De Abajo, F. J.; Kociak, M. Probing the photonic local density of states with electron energy loss spectroscopy. *Phys. Rev. Lett.* 2008, 100, 1−4.

(30) Wu, Y.; Li, G.; Camden, J. P. Probing Nanoparticle Plasmons with Electron Energy Loss Spectroscopy. *Chem. Rev.* 2018, 118, 2994−3031.

(31) Hage, F. S.; Kepaptsoglou, D. M.; Ramasse, Q. M.; Allen, L. J. Phonon Spectroscopy at Atomic Resolution. *Phys. Rev. Lett.* 2019, 122, 16103.

(32) Bellido, E. P.; Rossouw, D.; Botton, G. A. Toward 10 meV electron energy-loss spectroscopy resolution for plasmonics. *Microsc. Microanal.* 2014, 20, 767−778.

(33) Egerton, R. F. *Electron energy-loss spectroscopy in the electron microscope*, 3rd ed.; Springer Science & Business Media: New York, 2011;.

(34) Lagos, M. J.; Trügler, A.; Hohenester, U.; Batson, P. E. Mapping vibrational surface and bulk modes in a single nanocube. *Nature* 2017, 543, 529−532.

(35) Koh, A. L.; Bao, K.; Khan, I.; Smith, W. E.; Kothleitner, G.; Nordlander, P.; Maier, S. A.; Mccomb, D. W. Electron energy-loss spectroscopy (EELS) of surface plasmons in single silver nanoparticles and dimers: Influence of beam damage and mapping of dark modes. *ACS Nano* 2009, 3, 3015−3022.

(36) Isoniemi, T.; Maccaferri, N.; Ramasse, Q. M.; Strangi, G.; De Angelis, F. Electron Energy Loss Spectroscopy of Bright and Dark Modes in Hyperbolic Metamaterial Nanostructures. *Adv. Opt. Mater.* 2020, 8, 2000277.

(37) Wu, Y.; Li, G.; Cherqui, C.; Bigelow, N. W.; Thakkar, N.; Masiello, D. J.; Camden, J. P.; Rack, P. D. Electron Energy Loss Spectroscopy Study of the Full Plasmonic Spectrum of Self-Assembled Au-Ag Alloy Nanoparticles: Unraveling Size, Composition, and Substrate Effects. *ACS Photonics* 2016, 3, 130−138.

(38) Schoen, D. T.; Atre, A. C.; García-Etxarri, A.; Dionne, J. A.; Brongersma, M. L. Probing complex reflection coefficients in one-dimensional surface plasmon polariton waveguides and cavities using STEM EELS. *Nano Lett.* 2015, 15, 120−126.

(39) Song, F.; Wang, T.; Wang, X.; Xu, C.; He, L.; Wan, J.; Van Haesendonck, C.; Ringer, S. P.; Han, M.; Liu, Z.; et al. Visualizing plasmon coupling in closely spaced chains of Ag nanoparticles by electron energy-loss spectroscopy. *Small* 2010, 6, 446−451.

(40) Alexander, D. T. L.; Flauraud, V.; Demming-Janssen, F. Near-Field Mapping of Photonic Eigenmodes in Patterned Silicon Nanocavities by Electron Energy-Loss Spectroscopy. *ACS Nano* 2021, 15, 16501−16514.

(41) Raza, S.; Esfandyarpour, M.; Koh, A. L.; Mortensen, N. A.; Brongersma, M. L.; Bozhevolnyi, S. I. Electron energy-loss spectroscopy of branched gap plasmon resonators. *Nat. Commun.* 2016, 7, 1−10.

(42) Rossouw, D.; Couillard, M; Vickery, J.; Kumacheva, E.; Botton, G. A. Multipolar plasmonic resonances in silver nanowire antennas imaged with a subnanometer electron probe. *Nano Lett.* 2011, 11, 1499−1504.

(43) Nelayah, J.; Kociak, M.; Stéphan, O.; De Abajo, F. J. G.; Tencé, M.; Henrard, L.; Taverna, D.; Pastoriza-Santos, I.; Liz-Marzán, L. M.; Colliex, C. Mapping surface plasmons on a single metallic nanoparticle. *Nat. Phys.* 2007, 3, 348−353.

(44) Schaffer, B.; Hohenester, U.; Trügler, A.; Hofer, F. High-resolution surface plasmon imaging of gold nanoparticles by energy-filtered transmission electron microscopy. *Phys. Rev. B - Condens. Matter Mater. Phys.* 2009, 79, 1−4.

(45) Koh, A. L.; Fernández-Domínguez, A. I.; McComb, D. W.; Maier, S. A.; Yang, J. K. W. High-resolution mapping of electron-beam-excited plasmon modes in lithographically defined gold nanostructures. *Nano Lett.* 2011, 11, 1323−1330.

(46) Konečná, A.; Neuman, T.; Aizpurua, J.; Hillenbrand, R. Surface-Enhanced Molecular Electron Energy Loss Spectroscopy. *ACS Nano* **2018**, *12*, 4775−4786.

(47) Nicoletti, O.; De La Peña, F.; Leary, R. K.; Holland, D. J.; Ducati, C.; Midgley, P. A. Three-dimensional imaging of localized surface plasmon resonances of metal nanoparticles. *Nature* **2013**, *502*, 80−84.

(48) Hohenester, U. Simulating electron energy loss spectroscopy with the MNPBEM toolbox. *Comput. Phys. Commun.* **2014**, *185*, 1177−1187.

(49) Bernasconi, G. D.; Butet, J. my; Flauraud, V.; Alexander, D.; Brugger, J.; Martin, O. J. F. Where does energy go in electron energy loss spectroscopy of nanostructures? *ACS Photonics* **2017**, *4*, 156−164.

(50) García De Abajo, F. J. Optical excitations in electron microscopy. *Rev. Mod. Phys.* **2010**, *82*, 209−275.

(51) Schmidt, F. P.; Ditlbacher, H.; Hohenester, U.; Hohenau, A.; Hofer, F.; Krenn, J. R. Dark plasmonic breathing modes in silver nanodisks. *Nano Lett.* **2012**, *12*, 5780−5783.

(52) Krivanek, O. L.; Lovejoy, T. C.; Dellby, N.; Aoki, T.; Carpenter, R. W.; Rez, P.; Soignard, E.; Zhu, J.; Batson, P. E.; Lagos, M. J.; et al. Vibrational spectroscopy in the electron microscope. *Nature* **2014**, *514*, 209−212.

(53) Hachtel, J. A.; Lupini, A. R.; Idrobo, J. C. Exploring the capabilities of monochromated electron energy loss spectroscopy in the infrared regime. *Sci. Rep.* **2018**, *8*, 1−10.

(54) Lovejoy, T. C.; Corbin, G. C.; Dellby, N.; Hoffman, M. V.; Krivanek, O. L. Advances in Ultra-High Energy Resolution STEM-EELS. *Microsc. Microanal.* **2018**, *24*, 446−447.

(55) Gómez, D. E.; Teo, Z. Q.; Altissimo, M.; Davis, T. J.; Earl, S.; Roberts, A. The dark side of plasmonics. *Nano Lett.* **2013**, *13*, 3722−3728.

(56) Cuerda, J.; Rüting, F.; García-Vidal, F. J.; Bravo-Abad, J. Theory of lasing action in plasmonic crystals. *Phys. Rev. B - Condens. Matter Mater. Phys.* **2015**, *91*, 1−5.

(57) Hakala, T. K.; Rekola, H. T.; Väkeväinen, A. I.; Martikainen, J. P.; Nečada, M.; Moilanen, A. J.; Törmä, P. Lasing in dark and bright modes of a finite-sized plasmonic lattice. *Nat. Commun.* **2017**, *8*, 1−7.

(58) Omaghali, N. E. J.; Tkachenko, V.; Andreone, A.; Abbate, G. Optical sensing using dark mode excitation in an asymmetric dimer metamaterial. *Sensors (Switzerland)* **2014**, *14*, 272−282.

(59) Banerjee, S.; Amith, C. S.; Kumar, D.; Damarla, G.; Chaudhary, A. K.; Goel, S.; Pal, B. P.; Roy Chowdhury, D. Ultra-thin subwavelength film sensing through the excitation of dark modes in THz metasurfaces. *Opt. Commun.* **2019**, *453*, 124366.

(60) Zeng, Y.; Madsen, S. J.; Yankovich, A. B.; Olsson, E.; Sinclair, R. Comparative electron and photon excitation of localized surface plasmon resonance in lithographic gold arrays for enhanced Raman scattering. *Nanoscale* **2020**, *12*, 23768−23779.

(61) Mayerhöfer, T. G.; Popp, J. Periodic array-based substrates for surface-enhanced infrared spectroscopy. *Nanophotonics* **2018**, *7*, 39−79.

(62) Yoshimoto, D.; Saito, H.; Hata, S.; Fujiyoshi, Y.; Kurata, H. Characterization of Nonradiative Bloch Modes in a Plasmonic Triangular Lattice by Electron Energy-Loss Spectroscopy. *ACS Photonics* **2018**, *5*, 4476−4483.

(63) Park, S.; Mutz, N.; Schultz, T.; Blumstengel, S.; Han, A.; Aljarb, A.; Li, L. J.; List-Kratochvil, E. J. W.; Amsalem, P.; Koch, N. Direct determination of monolayer MoS2 and WSe2 exciton binding energies on insulating and metallic substrates. *2D Mater.* **2018**, *5*, 025003.

(64) Norcada. Available online: https://www.norcada.com/tech-info/ (accessed on Jan 10, 2023).

(65) Lumerical Inc. Available online: https://www.lumerical.com/ (accessed on Jan 10, 2023).

(66) Philipp, H. R. Optical Properties of Silicon Nitride. *J. Electrochem. Soc.* **1973**, *120*, 295.

(67) Johnson, P. B.; Christy, R. W. Optical Constant of the Nobel Metals. *Phys. Rev. B* **1972**, *6*, 4370−4379.